\documentclass[pre,twocolumn,showpacs]{revtex4}
\usepackage{graphics,graphicx,dcolumn,bm,fleqn,epic,eepic,float}
\usepackage{amssymb,amsmath,multirow,rotate,color}
\usepackage{subfigure}

\def \bJ {\mbox {\boldmath $J$}}
\def \bx {\mbox {\boldmath $x$}}
\def \bc {\mbox {\boldmath $c$}}
\def \bK {\mbox {\boldmath $K$}}

\begin{document}

\setcounter{page}{0}   
\input epsf

\title{\bf Localization transition in symmetric random matrices}
\author{F. L. \surname{Metz}, I. \surname{Neri} and D. \surname{Boll\'e}}
\affiliation{
Instituut voor Theoretische Fysica, Katholieke 
Universiteit Leuven, Celestijnenlaan 200D, B-3001
Leuven, Belgium }
\date{\today}
\thispagestyle{empty}

\begin{abstract}
We study the behaviour of the inverse participation ratio and the
localization transition in infinitely large random matrices
through the cavity method. Results are shown for two
ensembles of random matrices: Laplacian matrices on sparse 
random graphs and fully-connected L\'evy matrices. We derive 
a critical line separating localized from extended states 
in the case of L\'evy matrices. Comparison between 
theoretical results and diagonalization of finite random
matrices is shown.

\end{abstract}


\maketitle

\section{Introduction}

Since the pioneering work of Wigner in nuclear
physics \cite{W51}, random matrices have 
been intensively studied due to their wide range
of applications in several fields of physics and other 
disciplines. Some examples include
quantum chaos \cite{BGS84}, localization in electronic systems 
\cite{AAT73}, diffusion in random graphs \cite{BR88,F05}, 
finance \cite{Bouchaud1999,BP09} and complex
networks \cite{FDBV01}. 

In random matrix theory,
one is interested in physical quantities that can be
computed from the eigensolutions of a sample drawn
from an ensemble of $N \times N$ random matrices.
One of the main quantities of
interest is the density of states (DOS). In the case of the Gaussian 
orthogonal ensemble the DOS
obeys the Wigner semicircle law in the limit $N \rightarrow \infty$ \cite{Mehta}. There are
several ensembles in which the average DOS differs from
the semicircle law in a nontrivial way. 
The examples include
the ensemble of sparse matrices \cite{RB88,BG00,SC02,RCKT08,K08, RCKT09}, Laplacian 
matrices \cite{BM99,CGP99,D02,F05,AOI10} and L\'evy matrices \cite{CB94,AMA99,BJNPZ07}. 
Finite size effects do not play a
significant role and the DOS converges to its
large $N$ limit relatively fast. 

Another important quantity is the inverse participation 
ratio (IPR) since it provides valuable information about the nature
of the eigenstates. The IPR allows one to quantify the number of nonzero components in 
a certain eigenvector in the limit 
$N \rightarrow \infty$.
It is a suitable parameter to describe quantitatively
a delocalization-localization transition, since
it distinguishes between eigenstates that have a finite number
of nonzero components (localized states) and eigenvectors that
have an extensive number of nonzero components 
(extended states). The critical eigenvalue
separating localized from extended states is called the
mobility edge or the localization threshold.  
The mobility edge determines 
the Anderson transition in electronic systems \cite{AAT73}
and the emergence of particle traps in diffusion models 
on random lattices \cite{BR88,F05}.
An equation that determines the mobility edge 
in the ensemble of sparse random matrices was obtained
by means of the supersymmetric method \cite{FM91}. 
Within our knowledge there exists no full numerical
solution for the mobility edge by calculating the 
IPR with the supersymmetric method.

In contrast to the DOS, numerical diagonalization results show a 
significant dependence of the average IPR upon $N$ \cite{BM99,CGP99,K08}. 
Moreover, since localized states are usually present in the tails of the spectrum
of random matrices, one 
would have to diagonalize extremely large matrices 
to detect these states. 
In order to determine the localization threshold
in the limit $N \rightarrow \infty$, the imaginary part of the 
self-energy \cite{AAT73} and the variance of the density
of states \cite{EE92,CGMPV05} have been proposed as appropriate parameters.
However, a more quantitative description of the localization
transition would be obtained by calculating the average IPR
for infinitely large random matrices.

Laplacian matrices on sparse
random graphs \cite{BM99} and fully-connected 
L\'evy matrices \cite{CB94} are two examples of random matrices in which results
for the average IPR as
a function of the eigenvalue have been obtained only through 
diagonalization of finite matrices.
Laplacian random matrices arise, e. g., in the study
of diffusion on random graphs
\cite{BR88,D02,F05} and in the instantaneous normal modes approach
for liquid dynamics \cite{CGP99}.
L\'evy random matrices appear, e. g., in
models of spin-glasses with dipolar RKKY 
interactions \cite{Ciz1993}, in the study of
disordered electronic systems with interactions decaying as
a power-law of the distance \cite{CB94}, in portfolio
optimization \cite{Gal98} and in the study of correlations in data, for
instance, coming from
financial time series \cite{Burda06, Pol2010}.

In this work we calculate the eigenvalue-dependent IPR 
from the Green function corresponding to the
random matrix in the limit $N \rightarrow \infty$. The cavity
method provides a self-consistent equation for the Green function, which  
is easily solved through a population dynamics algorithm \cite{MP01}.
This approach is free of finite-size effects
and it can be used to study
the average IPR of different
ensembles of infinitely large random matrices. 
We present results for the average 
IPR as a function of the eigenvalue and the
presence of a localization transition for Laplacian matrices on a sparse
random graph and for fully-connected L\'evy matrices.
In the case of L\'evy matrices, we calculate a critical line that separates localized from extended 
eigenstates. The theoretical results for $N \rightarrow \infty$ are compared with numerical
diagonalization of finite matrices and with results of previous works.

In the next section we define the average quantities
of interest and we discuss how they can be calculated
in the limit $N \rightarrow \infty$. The results for 
the average IPR and the localization transition in the case of Laplacian and
L\'evy matrices are shown in section \ref{secresults}.
In section \ref{secconcl} we present our conclusions.
An equation that relates the Green function and the
IPR is derived in appendix \ref{sec:appB}.
We include a detailed discussion of the cavity method for the
ensemble of L\'evy matrices in 
appendix \ref{sec:appA}.


\section{The general setting} \label{sec2}

In this section we show how the DOS and the IPR 
are written in terms of the diagonal elements of the
Green function. The joint distribution of the 
real and imaginary parts of these diagonal elements
is the central quantity of interest to determine the average DOS and
IPR. We discuss how the cavity method can be
employed in order to calculate this distribution
in the large $N$ limit. 


\subsection{Random matrix parameters}

We define an ensemble $\Omega_{N}$ of symmetric
$N \times N$ random matrices with real elements. Assuming that a 
given matrix $\bJ \in \Omega_{N}$ has a set of
eigenvalues $\{\lambda_{\mu}\}_{\mu = 1,\dots,N}$
and normalized eigenvectors $\{ \mid\mu \rangle \}_{\mu = 1,\dots,N}$,
the DOS lying between $\lambda$ and $\lambda+d\lambda$ 
is given by
\begin{align}
\rho(\lambda) = \lim_{N \rightarrow \infty }  \frac{1}{N}\sum_{\mu=1}^{N} \delta(\lambda-\lambda_{\mu})\,\,.  \label{DOSdef} 
\end{align}

The IPR associated to a given eigenvector $\mid \mu \rangle$ is defined
as 
\begin{equation}
 Y^{N}_{\mu} = \sum_{i=1}^{N} (\psi^{i}_{\mu})^{4}\,\,,
\end{equation}
where 
$\psi^{i}_{\mu}= \langle i \mid \mu \rangle$ is the component
$i$ of the eigenvector $\mid \mu \rangle$. The set of
normalized vectors $\{ \mid i \rangle \}_{i = 1,\dots,N}$ 
is the canonical site basis. The IPR
allows one to distinguish between two
extreme situations in the limit $N \rightarrow \infty$. 
In a delocalized or extended region of the spectrum, a number of sites of $O(N)$
contributes to a given eigenstate. Due to the normalization
$\sum_{i=1} (\psi^{i}_{\mu})^2 =1$,
the components of a given $\mid \mu \rangle$ satisfy
$\psi^{i}_{\mu} = O(1/\sqrt{N})$ and, as
a consequence, we have $\lim_{N \rightarrow \infty} Y^{N}_{\mu} = 0$. 
In contrast, only a finite number of components $\{ \psi^{i}_{\mu} \}$
is nonzero for a localized eigenstate. 
The components of a
state $\mu$ localized on $d$ sites satisfy
$\psi^{i}_{\mu} = O(1/\sqrt{d})$
and the IPR is given by $Y^{N}_{\mu} = O(1/d)$ in the
limit $N \rightarrow \infty$. This simple analysis 
shows that the IPR is a suitable parameter for a quantitative description 
of the transition between extended and localized
eigenstates. 

Here we are interested in the average behaviour of
$Y^{N}_{\mu}$ over all the states in an infinitesimal 
region of the spectrum. We define the eigenvalue dependent IPR
\begin{align}
P(\lambda) = \lim_{N \rightarrow \infty }  \frac{1}{N \rho(\lambda)}  \sum_{\mu=1}^{N} \delta(\lambda-\lambda_{\mu}) Y^{N}_{\mu}\,\,, \label{IPRdef}
\end{align}
which is the limit $N\rightarrow \infty$ of the average value of
$Y^{N}_{\mu}$ over all the
states lying between $\lambda$ and $\lambda + d\lambda$ when $\rho(\lambda)$
is finite. 

The quantity that allows one to determine
the DOS and the IPR is
the Green function associated to $\bJ$. Its
diagonal elements are defined as follows
\begin{equation}
G^{N}_{i i} (z)  = (z-\bJ)_{i i}^{-1} = \sum_{\mu=1}^{N} \frac{(\psi^{i}_{\mu})^{2}}{z-\lambda_{\mu}} \,\,,
\label{Greendef}
\end{equation}
where $z$ is the complex variable $z= \lambda - i \epsilon$.
The DOS can be obtained in the limit $\epsilon \rightarrow 0^{+}$ according to
\begin{equation}
\rho(\lambda) =    \lim_{\epsilon \rightarrow 0^{+}} \lim_{N \rightarrow \infty }   \frac{1}{\pi N} 
\sum_{j=1}^{N} {\rm Im}G^{N}_{j j} (\lambda - i \epsilon)\,\,. 
\label{spectrGreen}
\end{equation}
The IPR is expressed by 
\begin{equation}
P(\lambda) = \lim_{\epsilon \rightarrow 0^{+}}  \lim_{N \rightarrow \infty } 
\frac{\epsilon}{\pi N \rho(\lambda)  } \sum_{j=1}^{N} |G^{N}_{j j} (\lambda - i \epsilon)|^{2} \,\,
\label{IPRGreen}
\end{equation}
in the regions of 
the spectrum where there are no degenerate
states. Equation (\ref{IPRGreen}) has been 
employed in the study of localization properties through the
supersymmetric approach \cite{FM91}.
In appendix \ref{sec:appB} we explain how
to derive eq. (\ref{IPRGreen}).
Finally, we rewrite eqs. (\ref{spectrGreen}) and 
(\ref{IPRGreen}) as follows
\begin{align}
\rho (\lambda) &=  \lim_{\epsilon \rightarrow 0^{+}} \frac{1}{\pi} \langle {\rm Im} \omega \rangle \,\,,
\label{spectrdistr}   \\
P(\lambda) &= \frac{1}{\pi \rho(\lambda)} \lim_{\epsilon \rightarrow 0^{+}} 
\epsilon \langle\, |\omega|^{2} \rangle \,\,,
\label{IPRdistr}
\end{align}
where we have introduced the average 
$\langle f(\omega) \rangle = \int d\omega W_{\lambda,\epsilon}(\omega) f(\omega)$ with 
respect to the joint distribution $W_{\lambda,\epsilon}(\omega)$
of the real and imaginary parts of $G^{N}_{i i}(z)$
\begin{equation}
W_{\lambda,\epsilon}(\omega) = \lim_{N \rightarrow \infty} \frac{1}{N} \sum_{j=1}^{N} 
\left \langle \delta \left[ \omega - G^{N}_{j j} \left( \lambda-i\epsilon \right) 
\right] \right \rangle_{\bJ} \,\,. 
\label{Wdef}
\end{equation}
We have assumed that the distribution of $G^{N}_{i i}(z)$
is a self-averaging quantity in the limit 
$N \rightarrow \infty$, with $\langle \dots \rangle_{\bJ}$ denoting the 
ensemble average with respect to the distribution of $\bJ$.
This implies that $\rho(\lambda)$ and 
$P(\lambda)$ are self-averaging quantities
in the limit $N \rightarrow \infty$.
The integrals present in $\langle f(\omega) \rangle$
run over the entire real and imaginary
axes of $\omega$. Our goal consists in calculating the 
distribution $W_{\lambda,\epsilon}(\omega)$, since
this quantity allows us to compute the average DOS 
and IPR through eqs. (\ref{spectrdistr}) and (\ref{IPRdistr}).


\subsection{A cavity calculation of the distribution of Green functions}

In this subsection we explain how the cavity method can be used
in order to calculate $W_{\lambda,\epsilon}(\omega)$. As an 
example, the technical details
involved in the calculations are shown in 
appendix \ref{sec:appA} for the case of fully-connected L\'evy matrices. 

The diagonal elements of the Green 
function can be written as a Gaussian integral
over the real vectors $\bx= (x_{1},\dots,x_{N})$, namely
\begin{equation}
G^{N}_{k k} (z)  = i \frac{\int d \bx \, x^{2}_{k} \, \exp{\left[-\frac{i}{2} \sum_{ij=1}^{N} x_i (z-\bJ)_{ij} x_j\right]} }
{\int d \bx \, \exp{\left[-\frac{i}{2} \sum_{ij=1}^{N} x_i (z-\bJ)_{ij} x_j \right]}} \,\,.
\label{GaussGreen}
\end{equation}
Defining the normalized complex function
\begin{equation}
\mathcal{P}_{N,z}(\bx) = \frac{\exp{\left[- H_{N,z}(\bx) \right] }}
{\int d \bx \, \exp{\left[- H_{N,z}(\bx)  \right]}} \,\,,
\label{measure}
\end{equation}
with
\begin{equation}
H_{N,z}(\bx) = \frac{i}{2} \sum_{ij=1}^{N} x_i (z-\bJ)_{ij} x_j \,\,,
\label{hamiltonian}
\end{equation}
the elements $G^{N}_{k k} (z)$ assume the form
\begin{equation}
G^{N}_{k k} (z)  = i \int d \bx \, x^{2}_{k}  \mathcal{P}_{N,z}(\bx) \,\, 
= i \int dx_k \, x^{2}_{k}  \mathcal{P}_{N,z}(x_k),
\label{Green1}
\end{equation}
where  $\mathcal{P}_{N,z}(x_k)$ follows from $\mathcal{P}_{N,z}(\bx)$ by
integrating over all variables besides $x_k$. More generally, we write for a
set $A$ of indices
\begin{eqnarray}
 \mathcal{P}_{N,z}\left(x_{A}\right) = \int \Big [ \prod_{i\notin A}dx_i \Big] 
\mathcal{P}_{N,z}\left(\bx\right). 
\end{eqnarray}
Equation (\ref{Green1}) shows that the Green functions $\{ G^{N}_{k k} (z) \}$ can be
computed for a single 
instance of $\bJ$ once we 
know how to calculate the local marginals 
$\{ \mathcal{P}_{N,z}(x_{k}) \}$.  

Disordered spin models on a random graph are defined 
through a probability measure on this graph. 
The cavity method provides an efficient way 
to compute the local marginals of disordered 
spin systems defined on fully-connected \cite{par1987} and 
finitely connected random graphs \cite{MP01}. The application of the cavity
method
to the study of random matrices relies on the
analogy between the random matrix problem and disordered
spin systems defined on graphs \cite{RCKT08}. Using this similarity 
one can apply the cavity
method to calculate the marginals $\{ \mathcal{P}_{N,z}(x_{k}) \}$. 

One can associate a random graph to a random matrix as follows: 
the graph contains $N$ nodes. A certain variable $x_{i}$ ($i=1,\dots,N$) 
is associated to the corresponding node $i$ of the graph. When $J_{ij}=0$ there is no edge between nodes
$i$ and $j$, while this pair of nodes is connected when $J_{ij} \neq 0$. The interaction strength between
nodes $i$ and $j$ is given by the value of $J_{ij}$, such that $\bJ$
specifies the topology of the random graph and the interaction strengths 
between the nodes. 

For a given graph instance $\mathcal{G}$, $\partial_{i}$  is the
the set of nodes
connected to a certain node $i$.  
The cavity method is based on the 
assumption that the distribution $\mathcal{P}_{N,z}(\bx_{\partial_{i}})$
of the variables in the neighbourhood $\partial_{i}$ factorizes on
the cavity graph $G^{(i)}$, i.e.
\begin{equation}
\mathcal{P}^{(i)}_{N,z}(\bx_{\partial_{i}}) = \prod_{j \in \partial_{i}}
\mathcal{P}^{(i)}_{N,z}(x_{j})\,\,.
\label{factassump}
\end{equation}
The cavity graph $G^{(i)}$ is the subgraph of
$G$ where node $i$ and
all its connections have been removed.
The marginals $\{ \mathcal{P}^{(i)}_{N,z}(x_{i}) \}$ are
defined on the cavity graph.
Since the local marginals $\{ \mathcal{P}_{N,z}(x_{j}) \}$ on the
real graph can be written in terms of the local marginals
on $\mathcal{G}^{(i)}$, one can solve first the problem on the cavity graph as a 
function of $\{ \mathcal{P}^{(i)}_{N,z}(x_{j}) \}$ and then
reconstruct the local marginals $\{ \mathcal{P}_{N,z}(x_{j}) \}$. 

In the case of disordered spin systems defined on finitely connected graphs, the cavity method is
also known as the Bethe-Peierls iterative method \cite{MP01}. In this
case, we expect that 
the factorization 
assumption (\ref{factassump}) holds outside the spin-glass phase 
since the graph looks locally like a tree.  When the graph 
is a tree the condition (\ref{factassump}) holds for $N \rightarrow \infty$.
In the case of disordered spin systems defined on fully-connected graphs, the
vanishing of the connected correlation functions for $N \rightarrow \infty$ ensures that the assumption
(\ref{factassump}) holds outside the spin-glass phase \cite{par1987}.

By employing the cavity method and following analogous calculations as done in
\cite{RCKT08}, we show in appendix \ref{sec:appA} that the 
marginal $\mathcal{P}_{N,z}(x_{k})$ at site $k$
is given by
\begin{equation}
\mathcal{P}_{N,z}(x_{k}) = \sqrt{\frac{i}{2 \pi G^{N}_{k k} (z)}}
\exp{\left( -\frac{i x^{2}_{k}}{2 G^{N}_{k k} (z) } \right)} \,\,.
\label{Gaussianmarg}
\end{equation}
The diagonal elements of the Green function 
are determined from the fixed-point solution of the equations
\begin{align}
G^{N,(k)}_{i i} (z)  &= \frac{1}{z - \sum_{j \in \partial_{i} 
\setminus k}  \, h^{N}_{i j}\big( J_{i j},G^{N,(i)}_{j j} (z)  \big) } \,\,, \label{caveqsR} \\
G^{N}_{i i} (z) &= \frac{1}{z - \sum_{j \in \partial_{i}} 
\, h^{N}_{ij} \big ( J_{i j},G^{N,(i)}_{j j} (z) \big) }\,\,, \label{realeqsR}
\end{align}
for $i=1,\dots,N$ and for all $k \in \partial_{i}$, where
$\partial_{i}$ is the set of all indices $j$ in a given
row $i$ such that $J_{ij} \neq 0$. The symbol
$\partial_{i} \setminus k$ denotes the set $\partial_{i}$
without index $k$. The quantities $\{ G^{N,(k)}_{ii}(z) \}$ are 
the diagonal elements of the Green function of the matrix following from the
original matrix through removal of
row $k$ and column $k$.
The specificity of the random matrices
under study can manifest itself only in the number of indices present
in $\partial_{i}$ and in the form
of the function $h^{N}_{i j} \big (  J_{i j}, G^{N,(i)}_{j j} (z) \big )$.

For fully-connected L\'evy matrices and Laplacian matrices considered in this
work, $h^{N}_{i j} \big (  J_{i j}, G^{N,(i)}_{j j} (z) \big )$
assumes the form:
\begin{itemize}
\item{L\'evy matrices: \\
\begin{equation}
h^{N}_{i j} \big (  J_{i j},  G^{N,(i)}_{j j} (z) \big ) = 
J^{2}_{i j}  G^{N,(i)}_{j j} (z)  \,\,,
\label{hijdefLM}
\end{equation}
}
\item{Laplacian matrices: \\
\begin{equation}
h^{N}_{i j} \big (  J_{i j},  G^{N,(i)}_{j j} (z) \big ) = 
\frac{J^{2}_{i j} G^{N,(i)}_{j j} (z) }{1+ J_{i j}  G^{N,(i)}_{j j} (z)}
- J_{i j} \,\,.
\end{equation}
}
\end{itemize}
Equations (\ref{caveqsR}) and (\ref{realeqsR})
with $h^{N}_{i j} \big (  J_{i j}, G^{N,(i)}_{j j} (z)  \big )$
given by (\ref{hijdefLM})
have been obtained previously in the study
of fully-connected L\'evy matrices \cite{CB94}
and sparse random matrices \cite{RCKT08}.  

A self-consistent equation for $W_{\lambda,\epsilon}(\omega)$ is obtained
by substituting eq. (\ref{realeqsR}) in eq. (\ref{Wdef}) and performing
the average over the ensemble of random matrices. We have solved numerically 
this self-consistent equation through a population dynamics 
algorithm \cite{MP01}, which consists in parametrizing the
distribution $W_{\lambda,\epsilon}(\omega)$ by a large population 
of stochastic variables representing instances of $\omega$. At 
each iteration step, one of these variables
is chosen at random and updated according to its probability 
distribution, until a stationary form for $W_{\lambda,\epsilon}(\omega)$ 
is reached. 
A detailed discussion of the population dynamics method
in the context of random matrices and the corresponding algorithm
are presented in \cite{K08}.

According to eqs. (\ref{spectrdistr}) and (\ref{IPRdistr}), one
has to obtain numerical results for the distribution
$W_{\lambda,\epsilon}(\omega)$ in the
limit $\epsilon \rightarrow 0$. One has to calculate 
$W_{\lambda,\epsilon}(\omega)$ for very small but finite values of 
$\epsilon$, since Dirac delta peaks 
might arise, for instance, in the spectrum of sparse random matrices.
In this way the Dirac delta peaks
are approximated by Lorentzian functions with a finite
width $\epsilon$ \cite{RCKT08,K08}.
In the next section, we specify the 
ensembles of random matrices and the corresponding
distribution $W_{\lambda,\epsilon}(\omega)$
for each case.


\section{Results} \label{secresults}

In this section we show the results for 
two different ensembles of symmetric random matrices:
Laplacian matrices on sparse random graphs and fully-connected
L\'evy matrices.
In both cases we focus on the behaviour of the
average IPR and the presence of a localization
transition. 


\subsection{Laplacian matrices}

The elements of the Laplacian matrix $\bJ$
on a random graph can be defined according to \cite{K08}
\begin{equation}
J_{i j} = c_{i j} K_{i j} - \delta_{i j} \sum_{k=1}^{N} c_{i k} K_{i k} \,\,,
\label{defLapl}
\end{equation}
in which $\bc$ and $\bK$ are symmetric matrices.
We consider here only the case in which the elements
of the connectivity matrix $\bc$ are i.i.d.r.v drawn from the distribution
\begin{equation}
p_{c}(c_{i j}) = \left( 1 -\frac{c}{N} \right) \delta_{c_{i j},0} + \frac{c}{N} \delta_{c_{i j},1} \,\,,
\end{equation}
with $c_{i i } = 0$ for $\forall \, i$. In the limit $N \rightarrow \infty$, $\bJ$ has a sparse structure 
and the number of nonzero elements per row exhibits a Poissonian
distribution with average $c$. The nonzero elements $\{ K_{ij} \}$ are drawn
according to the distribution $p_{K}(K_{ij})$. We consider here 
two different
cases: (i) the elements $\{ K_{ij} \}$
assume a fixed value $K_{ij} = -1/c$, such that $p_{K}(K_{ij}) = \delta(K_{ij}+1/c)$; (ii) 
the elements $\{ K_{ij} \}$
are drawn from a Gaussian distribution with
zero mean and variance $1/c$. 

Inserting eq. (\ref{realeqsR}) in eq. (\ref{Wdef}) and performing the
ensemble average, one can derive the
following self-consistent equation for $W_{\lambda,\epsilon}(\omega)$
\begin{align}
&W_{\lambda,\epsilon}(\omega) = \sum_{k=0}^{\infty} \frac{e^{-c} \, c^{k}}{k!}
\int \left[ \prod_{l=1}^{k} d \omega_{l} W_{\lambda,\epsilon}(\omega_{l}) \right] \nonumber \\
&\times \int \left[ \prod_{l=1}^{k} d K_{l} \, p_{K}(K_{l}) \right]
\delta \left(\omega - \frac{1}{z - \sum_{l=1}^{k}  H(\omega_{l},K_{l})} \right) \,\,. 
\label{eqW}
\end{align}
The updating 
function $H(\omega,K)$ 
is given by
\begin{equation}
H(\omega,K) = \frac{K^{2} \omega}{1+ K \omega}
- K \,\,.
\end{equation}
The population dynamics algorithm 
can be used to solve eq. (\ref{eqW}) numerically \cite{MP01}.

\begin{figure}[h]
\centering
\includegraphics[width=8.5cm,height=6.0cm]{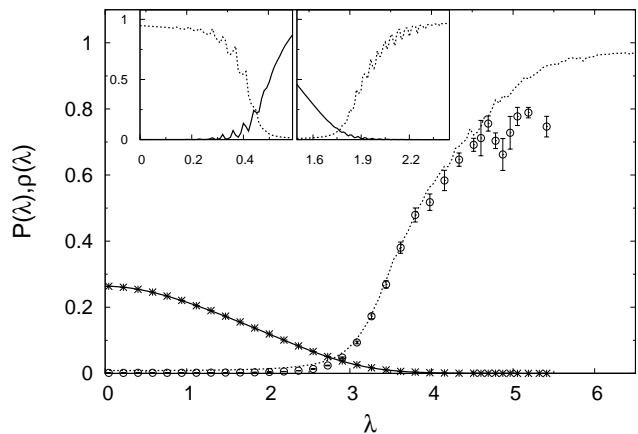}
\caption{Population dynamics results for the average DOS (solid lines) 
and IPR (dotted lines) for Laplacian matrices
with $c=20$ and elements $\{ K_{l} \}$ drawn from a Gaussian
distribution. These results were obtained with $\epsilon = 0.001$ and
a population of
$5 \times 10^{6}$ samples for the distribution $W_{\lambda,\epsilon}(\omega)$.
Numerical diagonalization results for the DOS ($\ast$) and the IPR ($\circ$)
obtained with an ensemble of $500$ matrices of dimension $N=3000$ are shown. 
Error bars for the IPR are indicated.
The insets show population dynamics results for the
DOS (solid lines) and the IPR (dotted lines) as a function of $\lambda$ in the left
and right tails of 
the spectrum of Laplacian matrices with $c=20$, $\epsilon = 0.001$ and 
fixed elements $\{ K_{l} \}$.
}
\label{fig1}
\end{figure}

In fig. \ref{fig1} we illustrate the results for $\rho(\lambda)$
and $P(\lambda)$ obtained from the
population dynamics algorithm and from diagonalization
of finite matrices for $c=20$. The spectra
of random matrices with a sparse structure contain delta peaks 
located at the eigenvalues of isolated finite-size
clusters for any value of $c$ \cite{BG00,K08,RCKT08}.
Eq. (\ref{IPRGreen}) gives an approximation for the IPR 
in these regions of the 
spectrum due to the presence of degenerate states.
In order to minimize the effect of these
singular contributions, we have chosen a large 
value of $c$. The main graph of fig. \ref{fig1} illustrates
$\rho(\lambda)$ and $P(\lambda)$ for the Gaussian 
distribution of $\{ K_{l} \}$ only when $\lambda \geq 0$, since
the spectrum is symmetric around zero. The insets
show the behaviour of $\rho(\lambda)$ and $P(\lambda)$
in the tails of the spectrum
for the case of fixed $K_{l} = - 1/c$. In both cases, $P(\lambda)$ is vanishingly small
in the central part of the spectrum, corresponding to 
a region of extended eigenstates.
The eigenvectors undergo a localization transition in the tails
of the spectrum, as shown by the increase
of $P(\lambda)$. It is difficult to determine the
IPR through numerical diagonalization in the tails
of the spectrum since one has to diagonalize
extremely large matrices.

By means of numerical diagonalization and a single defect
approximation (SDA) \cite{BM99,SC02}, the authors of 
\cite{BM99} have studied the eigenstates corresponding to the 
regular peaks that appear for large and small eigenvalues in
the spectrum of Laplacian matrices with fixed values of $\{ K_{l} \}$. 
They have found that these states are 
localized on a finite number of sites that have a small
or large connectivity in comparison to the mean $c$.
Fig. \ref{fig1} complements these results by showing
that $P(\lambda) \rightarrow 1$ for large values of $\lambda$, which
means that eigenstates corresponding to large eigenvalues 
are localized on a single site. The presence of the peaks is reduced 
when one introduces Gaussian disorder in the elements $\{ K_{l} \}$. 

The agreement between
diagonalization and theoretical results for $\rho(\lambda)$ in fig. \ref{fig1} is 
very good. We have found that $\rho(\lambda)$ 
depends weakly on $N$ or $\epsilon$ in the case of numerical
diagonalization or population dynamics, respectively.
In the case of $P(\lambda)$, both results
exhibit a very good agreement in the central 
region of the spectrum and in parts of the tails,
for the particular values of $\epsilon$ and $N$ chosen. 
The results show a discrepancy in the far regions of the
tails where the eigenvalues are rare, as shown by
the higher fluctuations in the numerical diagonalization, 
illustrated by the error bars.
However, the average IPR
depends on the values of $N$ and 
$\epsilon$.
\begin{figure}[h]
\centering
\includegraphics[width=8.0 cm,height=9.5cm]{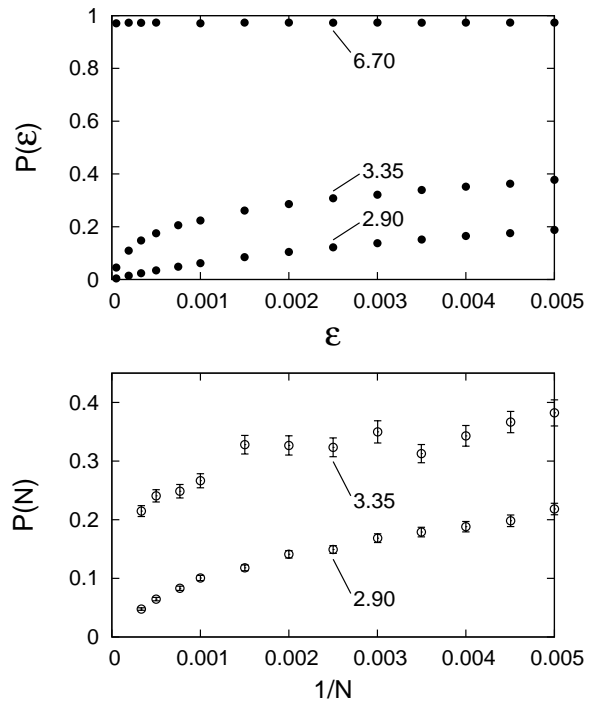}
\caption{Average IPR for Laplacian matrices as
a function of $\epsilon$ (population dynamics, top graph) and $N$ (numerical 
diagonalization, bottom graph) for $c=20$ and different 
numerical values of $\lambda$, which are shown explicitly on the figure. The elements $\{ K_{l} \}$ are drawn from the 
Gaussian distribution described in the text. The diagonalization results were obtained considering an ensemble
of $500$ matrices for each $N$, with error bars included in this case. The population dynamics 
results were obtained with a population of $5 \times 10^{6}$ samples.
}
\label{fig2}
\end{figure}

Figure \ref{fig2} illustrates the behaviour of $P(\lambda)$ as a 
function of $N$ and $\epsilon$ for the Laplacian matrix
with Gaussian elements $\{ K_{l} \}$ and $c=20$. 
The results show that, for $\lambda = 2.90$ and 
$\lambda=3.35$, the average IPR goes to zero when 
$\epsilon \rightarrow 0$.
Accordingly, the diagonalization results for $P(\lambda)$ exhibit 
a similar qualitative behaviour as $N$ increases. 
For $\lambda=6.70$, the
average IPR has a finite value for $\epsilon \rightarrow 0$, 
since this is the region of localized eigenstates.
These results indicate that the localization transition
presented in fig. \ref{fig1} becomes sharper for $N \rightarrow \infty$.
By employing numerical diagonalization methods, the authors of \cite{BM99}
have found the value $\lambda_{c} \simeq 1.67$ for the mobility edge
on the right tail in the case of Laplacian matrices
with $c=20$ and fixed elements $\{ K_{l} \}$. 
We have calculated
approximately the localization threshold $\lambda_{c}$ on the right tail by using
the $\epsilon$ independence of $P(\lambda)$ as a criterion to determine 
the localized region.
For $c=20$ we have found the values $\lambda_{c} \simeq 1.95$ and 
$\lambda_{c} \simeq 5.65$ for Laplacian matrices with fixed 
and Gaussian elements $\{ K_{l} \}$, respectively.


\subsection{L\'evy matrices} \label{seclevy}

The fully-connected L\'evy matrix is a symmetric matrix in which 
$J_{ii}=0$ for $\forall \, i$. The nondiagonal elements are i.i.d.r.v.~drawn
from the L\'evy distribution $P_{\alpha}(J)$, defined
through the characteristic function $L_{\alpha}(q)$
\begin{eqnarray}
 P_{\alpha}(J) \equiv \int \frac{dq}{2\pi}\exp\left(-iqJ\right)L_{\alpha}(q)
\label{defStable}.
\end{eqnarray}
The characteristic function is of the form
\begin{eqnarray}
\ln L_{\alpha}(q)  =  - \left|\frac{q}{\sqrt{2} N^{1/\alpha}}\right|^{\alpha} \,.\,
\label{defChar}
\end{eqnarray}
The distribution $P_{\alpha}(J)$ is fully
determined by the parameter $\alpha \in (0,2]$.
For $\alpha <2$, $\alpha$ characterizes the power-law decay of 
$P_{\alpha}(J)$.
We consider only L\'evy
distributions centered around zero.
The scaling with 
$N$ in eq. (\ref{defChar}) ensures that the spectrum
converges to a stable form in the limit $N \rightarrow \infty$ \cite{CB94}. 
The distribution of
Green functions does not depend
on the skewness parameter \cite{CB94,BJNPZ07}.

For $\alpha=2$ we recover the Gaussian orthogonal ensemble
since $P_{\alpha}(J)$ is a Gaussian distribution
with zero mean and variance $1/N$.
For $\alpha < 2$, the asymptotic behaviour 
of $P_{\alpha}(J)$ for $|J| \rightarrow \infty$ can
be derived from the explicit form of $L_{\alpha}(q)$:
\begin{equation}
\lim_{|J| \rightarrow \infty}  P_{\alpha}(J) = 
 \frac{C_{\alpha}}{N |J|^{\alpha+1}} ,
\label{asympP}
\end{equation}
where
\begin{equation}
C_{\alpha} = \left( \frac{1}{\sqrt{2}} \right)^{\alpha}\frac{1}{\pi} 
\sin\left(\frac{\alpha \pi}{2}\right)\Gamma(\alpha+1) . 
\label{Cdef}
\end{equation}
The integrals for the second and higher moments of
the distribution diverge for $\alpha < 2$ due to the 
power-law decay illustrated by eq.~(\ref{asympP}).

Due to the power-law tails of the L\'evy distribution, each
row of $\bJ$ contains an infinite number of
elements of order $O(N^{-1/\alpha})$ and a finite number
of elements of order $O(1)$. For small values of $\alpha$, it has
been argued that fully-connected L\'evy matrices can be
seen as sparse random matrices \cite{CB94,AMA99}.
The spectrum of L\'evy matrices has been calculated 
with the cavity method in previous works \cite{CB94,BJNPZ07} 
and an equation for 
the distribution of Green functions has been determined using the generalized central
limit theorem \cite{Gn1954}.
We have followed a different approach
to calculate the distribution of Green functions, 
in which the underlying sparse character of L\'evy matrices
becomes transparent. Besides that, the resulting self-consistent
equation can
be solved through a population dynamics algorithm, which
is a practical advantage in comparison with previous works
where one has to deal with a complicated 
system of integral equations \cite{BJNPZ07}. 

In order to take the ensemble average and the limit $N\rightarrow \infty$ of
the distribution of Green functions, we introduce a cutoff $\gamma$ that makes an 
explicit distinction between strong matrix elements $J_{ij}> \gamma$ and   
weak matrix elements $J_{ij}< \gamma$ (see appendix \ref{sec:appA}).  This
trick has been introduced in spin systems in \cite{NMB10, Jan10}
The backbone of strong matrix elements 
can be treated as a sparse random matrix, leading to the
following self-consistent equation
\begin{align}
&W_{\lambda,\epsilon, \gamma}(\omega) = \sum_{k=0}^{\infty} \frac{e^{-c_{\gamma}} \, c_{\gamma}^{k}}{k!}
\int \left[ \prod_{l=1}^{k} d \omega_{l} W_{\lambda,\epsilon, \gamma}(\omega_{l}) \right] \nonumber \\
&\times \int \left[ \prod_{l=1}^{k} d J_{l} \, p_{J, \gamma}(J_{l}) \right] \nonumber \\
&\times \delta \left(\omega - \frac{1}{z - \sigma_{\gamma}^{2} 
\langle \omega \rangle -  \sum_{l=1}^{k} H(\omega_{l},J_{l})} \right) \,\,,
\label{eqWLevy}
\end{align}
where
\begin{align}
H(\omega,J) &= J^{2} \omega   \\
p_{J,\gamma}(J) &= \left\{ \begin{array}{ccc}\frac{\alpha \gamma^{\alpha}}{2 |J|^{\alpha+1}}&  &|J|>\gamma \\ 0& &|J|<\gamma  \end{array}
\right. \,\,,
\end{align}
and 
\begin{align}
c_{\gamma} &= \frac{2 C_{\alpha}}{\alpha \gamma^{\alpha}} \,\,, \\
\sigma_{\gamma}^{2}&=\frac{2 \gamma^{2-\alpha} C_{\alpha}}{2-\alpha}\,\,.
\end{align}
The distribution of Green functions follows from $W_{\lambda, \epsilon}(\omega)=
\lim_{\gamma\rightarrow 0}W_{\lambda,\epsilon,
\gamma}(\omega)$.

The quantity $c_{\gamma}$
is the average number of strong matrix elements
per row and $p_{J,\gamma}(J)$ denotes their distribution. 
The contribution of the infinite number of weak matrix elements
is taken into account through the law of large numbers, leading
to a term proportional to their variance
$\sigma_{\gamma}^{2}$. Eq. (\ref{eqWLevy}) shows that the 
distribution $W_{\lambda,\epsilon, \gamma}(\omega)$ contains
a part coming from a sparse random matrix of strong 
matrix elements and an average contribution due
to the weak matrix elements. 

\begin{figure}[H]
\centering
\includegraphics[width=7.3cm,height=7.0cm]{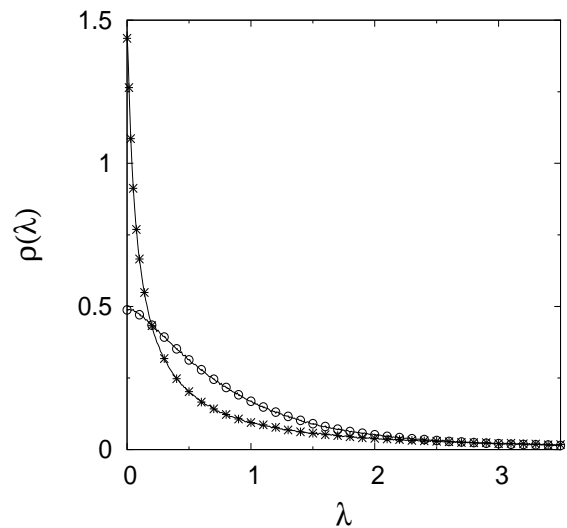}
\caption{
Comparison between numerical diagonalization (full lines) and population
dynamics results (different types of symbols) for the average DOS of L\'evy matrices
with $\alpha=0.75$ ($\ast$) and $\alpha=1.25$ ($\circ$). The diagonalization
results were obtained considering an ensemble of $1000$ matrices
of dimension $N=1500$. The population dynamics results
were obtained considering $\epsilon = 0.001$, $\gamma=0.01$ and
a population of $10^{6}$ samples.
}
\label{fig3}
\end{figure}

We have solved eq. (\ref{eqWLevy}) through a population
dynamics algorithm. The idea is to obtain results for small
values of the cutoff $\gamma$. In fig. \ref{fig3} we show results for the 
DOS of L\'evy matrices obtained from the numerical diagonalization of finite 
matrices and from the numerical solution of eq. (\ref{eqWLevy}). 
The DOS of L\'evy matrices is symmetric around zero.
For both values of $\alpha$ the agreement between diagonalization
and population dynamics results is excellent.
\begin{figure}[h]
\centering
\includegraphics[width=7.3cm,height=7.0cm]{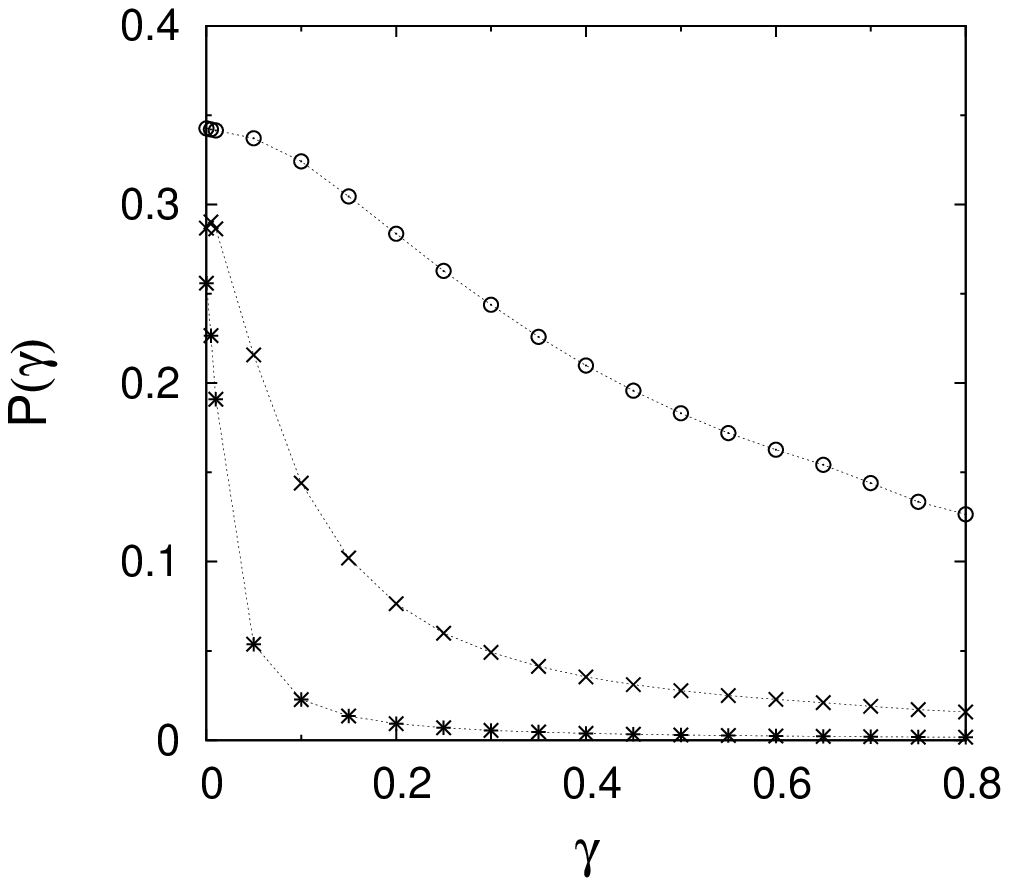}
\caption{Population dynamics results for the average IPR
of L\'evy matrices as
a function of $\gamma$ for $\alpha=0.5$, $\lambda=1$ and a 
population of $5 \times 10^{6}$ samples. Three different values of $\epsilon$
are shown: $\epsilon = 0.01$ ($\circ$), $\epsilon = 0.001$ ($\times$) 
and  $\epsilon = 0.0001$ ($\ast$). The average IPR depends 
on $\epsilon$ for $\gamma \rightarrow 0$.
}
\label{fig4}
\end{figure}

For $\alpha \rightarrow 2$, we obtain
$c_{\gamma} \rightarrow 0$ and $\sigma_{\gamma}^{2} \rightarrow 1$, and
the distribution $W_{\lambda,\epsilon, \gamma}(\omega)$ reduces to
the simple form
\begin{equation}
W_{\lambda,\epsilon}(\omega) = \delta \left( \omega - \frac{1}{z-\langle \omega \rangle }  \right)
\label{Wigdelt}
\,\,.
\end{equation}
When inserted in the definition of $\langle \omega \rangle$, eq. (\ref{Wigdelt}) 
gives rise to a quadratic equation
for $\langle \omega \rangle$ whose solution leads
to the Wigner semicircle law by means of eq. (\ref{spectrdistr}).

\begin{figure}[h]
\centering
\includegraphics[width=7.3cm,height=7.0cm]{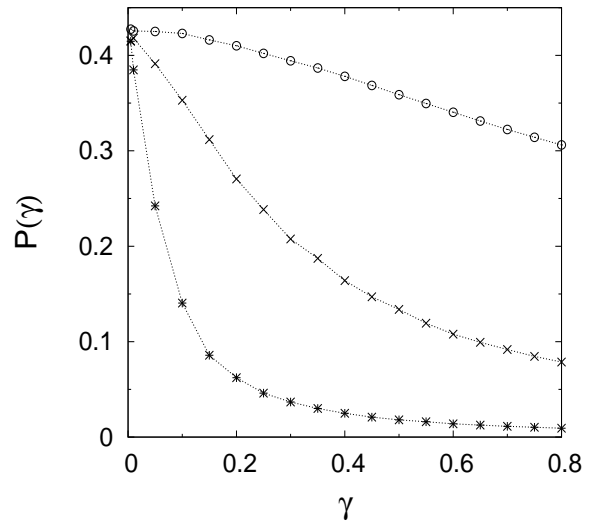}
\caption{Population dynamics results for the average IPR of
L\'evy matrices as
a function of $\gamma$ for $\alpha=0.5$, $\lambda=5$ and
a population of $5 \times 10^{6}$ samples. Three different values of $\epsilon$
are shown: $\epsilon = 0.01$ ($\circ$), $\epsilon = 0.001$ ($\times$) 
and  $\epsilon = 0.0001$ ($\ast$). The average IPR is
independent of $\epsilon$ for $\gamma \rightarrow 0$.
}
\label{fig5}
\end{figure}

In figs. \ref{fig4} and \ref{fig5} we illustrate the population
dynamics results for the
behaviour of $P(\gamma)$ as a function of $\gamma$
for $\lambda =1$ and $\lambda =5$, respectively. In both
figures we consider $\alpha = 0.5$ and three different
values of $\epsilon$. Figure \ref{fig4} shows that for
$\gamma \rightarrow 0$ the average IPR decreases 
for decreasing values of $\epsilon$, which is an indication
that $\lambda=1$ corresponds to a region
with delocalized states. As
$\gamma$ decreases in fig. \ref{fig5}, $P(\gamma)$ 
converges to a finite value that does not
depend on $\epsilon$, which corresponds
to a region of localized states. 

\begin{figure}[h]
\centering
\includegraphics[width=7.3cm,height=7.0cm]{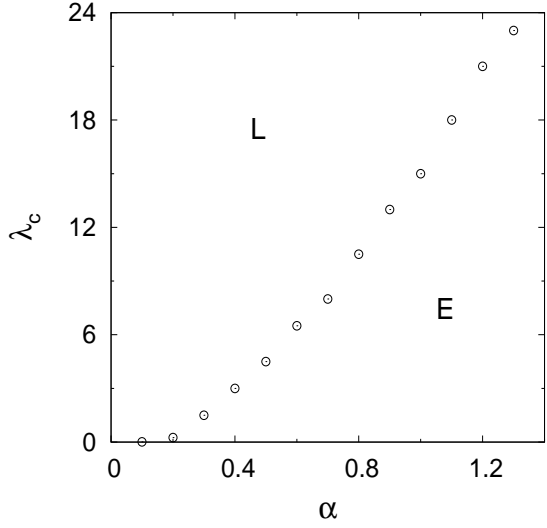}
\caption{Population dynamics results for the critical line 
separating localized (L) from extended
states (E) in the fully-connected L\'evy matrix. The results
were obtained considering $\epsilon = 0.0001$, $\gamma = 0.008$ and
a population of $10^{6}$ samples.
}
\label{fig6}
\end{figure}

We have used the $\epsilon$ independence of $P(\lambda)$ 
in the localized region as a criterion to calculate approximately 
the localization
threshold in the limit $N \rightarrow \infty$.
In fig. \ref{fig6} we present 
results for the critical line separating localized from
extended states in the $(\alpha,\lambda)$ plane for $\gamma=0.008$. 
For small values of $\alpha$, $c_{\gamma}$ is small and the tails of the distribution $p_{J,\gamma}(J)$ are very 
long. In this case the sparse matrix character of L\'evy matrices is 
highlighted and the region of localized eigenstates is larger.
For increasing values of $\alpha$, the parameter $c_{\gamma}$ increases 
and the tails of the distribution $p_{J,\gamma}(J)$ are less long.
The distinction between strong and weak matrix elements becomes
less important and the fully-connected character of L\'evy matrices is 
highlighted, leading to a larger region of extended eigenstates.
The results for the localization transition when $\alpha$ is
large are
very noisy due to the larger values of $\lambda_{c}$ involved in the
calculations. This makes the results for $\alpha > 1.3$ very
inaccurate. Besides that, the population dynamics
algorithm becomes slower for increasing values of $c_{\gamma}$. 
We have obtained numerically that the average
number of strong matrix elements $c_{\gamma}$ reaches its
maximum value at $\alpha \simeq 2$ when $\gamma \rightarrow 0$. This indicates
that the region of extended states is the largest possible for
$\alpha \simeq 2$.

For $\alpha=0.5$ the population dynamics results show that $P(\lambda) \rightarrow 1/2$ 
as $\lambda \rightarrow \infty$. This result agrees with 
the discussion presented in \cite{CB94}. According to 
this work, due to the strong fluctuations of the 
L\'evy matrix elements, the 
eigenstates corresponding to large eigenvalues are localized
on pairs of very strongly interacting sites, which
leads to an IPR equal to $1/2$.
Based mostly on numerical diagonalization 
results \cite{CB94,AMA99}, previous works point to the 
presence of three regions: a region of extended eigenstates,
a strictly localized region, with exponentially
localized eigenstates, and 
a mixed region, exhibiting both localized and
extended features. The results of the literature \cite{CB94} 
suggest that the eigenstates decay algebraically
in the mixed region. From the study of the IPR
one can not distinguish between these two different 
types of localized states. 

%


\section{Conclusion} \label{secconcl}

In this paper we have studied the localization of eigenvectors 
of random matrices through the calculation of the inverse participation 
ratio with the cavity method. We have found a self-consistent equation for the inverse participation ratio 
in the limit $N\rightarrow \infty$, which can be solved numerically 
through a population dynamics algorithm. Therefore, this approach contains no finite size effects 
in contrast with numerical diagonalization methods. The resultant equations for the inverse 
participation ratio are conjectured to be exact for Laplacian matrices on sparse
random graphs and for fully connected 
L\'evy matrices in the limit $N\rightarrow \infty$.  

We have calculated the inverse participation ratio of Laplacian 
matrices on sparse random graphs. The spectrum is characterized by a delocalized part 
centered around zero and a localized part in the edges of the spectrum. 
The states corresponding to large eigenvalues are 
localized on single sites with large degrees in comparison
to the average connectivity \cite{BM99}. 
Numerical diagonalization results for the inverse participation ratio appear to converge to the corresponding 
theoretical values for $N\rightarrow \infty$ when there is no degeneracy in the eigenvalues.  

The matrix elements of random L\'evy matrices are
drawn from a distribution with power-law tails characterized by an exponent 
$\alpha$. In previous works \cite{CB94, AMA99}, the localization properties of
L\'evy matrices have been studied using mainly diagonalization results.
These results indicate that the eigenstates of L\'evy matrices undergo a transition from a 
delocalized to a localized phase.
However, large finite size effects are present in such calculations. Using the cavity approach we 
have determined a transition line separating localized from delocalized 
states in the $(\alpha, \lambda)$ plane by studying the behaviour
of the inverse participation ratio in the limit 
$N\rightarrow \infty$. Our results confirm the presence of a region with localized 
states for large eigenvalues, where the states are localized on pairs
of very strongly interacting sites \cite{CB94}.

While we were writing this paper a preprint appeared on the arxiv addressing 
similar issues \cite{Biroli}. The authors of this paper have determined the location 
of the Anderson transition in electronic systems on Bethe lattices using the cavity 
method, while we have focused on the localization properties of random matrix ensembles. 


\appendix


\section{The eigenvalue-dependent IPR and the Green function}  \label{sec:appB}

In this appendix we show how the eigenvalue-dependent IPR
can be expressed in terms of $G^{N}_{i i} (\lambda - i \epsilon)$
by means of eq. (\ref{IPRGreen}). By substituting eq. (\ref{Greendef}),
one can write down the following quantity
\begin{equation}
\mathcal{G}^{N}_j(\lambda) = \lim_{\epsilon \rightarrow 0} \epsilon |G^{N}_{j j}(\lambda - i \epsilon)|^2\,\,,
\end{equation}
in the form
\begin{align}
\mathcal{G}^{N}_j(\lambda) = \lim_{\epsilon \rightarrow 0}  \epsilon
 \sum_{\mu=1}^{N} \frac{(\psi_{\mu}^{j})^{4}}{(\lambda - \lambda_{\mu})^2 + \epsilon^2}
+ \lim_{\epsilon \rightarrow 0}  D^{N}_j(\epsilon,\lambda)\,\,,
\label{invPapp}
\end{align}
where we have defined the non-diagonal contribution
\begin{equation}
D^{N}_j(\epsilon,\lambda) =  \epsilon \sum_{\mu=1}^{N}  \frac{(\psi_{\mu}^{j})^{2}}{\lambda - \lambda_{\mu} + i \epsilon}
\sum_{\nu \neq \mu}  \frac{(\psi_{\nu}^{j})^{2}}{\lambda - \lambda_{\nu} - i \epsilon} \,\,.
\end{equation}

Assuming there is a set of eigenvalues $\mathcal{A} = \{ \lambda_{k_1},\dots, \lambda_{k_K}   \}$ such
that $\lambda = \lambda_{k_i}$ for any $\lambda_{k_i} \in \mathcal{A}$, we 
can rewrite $D^{N}_l(\epsilon,\lambda)$ as follows
\begin{align}
&D^{N}_l(\epsilon,\lambda) = \frac{1}{i} \sum_{j=1}^{K} (\psi_{k_{j}}^{l})^{2} \sum_{\nu \neq k_{j}}
\frac{(\psi_{\nu}^{l})^{2}}{\lambda - \lambda_{\nu} - i \epsilon} \nonumber \\
&-
\frac{1}{i} \sum_{j=1}^{K} (\psi_{k_{j}}^{l})^{2} \sum_{\nu \neq \mathcal{A}}
\frac{(\psi_{\nu}^{l})^{2}}{\lambda - \lambda_{\nu} + i \epsilon}  \nonumber \\
&+ \epsilon \sum_{\mu \neq \mathcal{A}} \frac{(\psi_{\mu}^{l})^{2}}{\lambda - \lambda_{\mu} + i \epsilon}
\sum_{\nu \neq \mu, \mathcal{A}}
\frac{(\psi_{\nu}^{l})^{2}}{\lambda - \lambda_{\nu} - i \epsilon}\,\,.
\end{align}
In the absence of degenerate states in the spectrum, the 
set $\mathcal{A}$ is simply given by 
$\mathcal{A} = \lambda_{k}$, which reads
\begin{align}
&D^{N}_j(\epsilon,\lambda) = \frac{1}{i}  (\psi_{k}^{j})^{2} \sum_{\nu \neq k}
\frac{(\psi_{\nu}^{j})^{2}}{\lambda - \lambda_{\nu} - i \epsilon} \nonumber \\
&-
\frac{1}{i}  (\psi_{k}^{j})^{2} \sum_{\nu \neq k}
\frac{(\psi_{\nu}^{j})^{2}}{\lambda - \lambda_{\nu} + i \epsilon}  \nonumber \\
&+ \epsilon \sum_{\mu \neq k} \frac{(\psi_{\mu}^{j})^{2}}{\lambda - \lambda_{\mu} + i \epsilon}
\sum_{\nu \neq \mu, k}
\frac{(\psi_{\nu}^{j})^{2}}{\lambda - \lambda_{\nu} - i \epsilon}\,\,.
\end{align}
Thus we obtain $\lim_{\epsilon \rightarrow 0} D^{N}_i(\epsilon,\lambda) = 0$
and eq. (\ref{invPapp}) assumes the form
\begin{align}
 \mathcal{G}^{N}_i(\lambda) = \pi 
\sum_{\mu=1}^{N} (\psi_{\mu}^{i})^{4}\delta(\lambda - \lambda_{\mu})\,\,.
\label{invPappaux}
\end{align}
By summing the above equation over all the sites
and dividing by $N \rho(\lambda)$ we 
obtain, in the limit $N \rightarrow \infty$, the
identity that relates the eigenvalue-dependent IPR with
$G^{N}_{ii}(\lambda - i \epsilon)$ (see eq. (\ref{IPRGreen})).
The IPR associated to the state $\mu$ is defined as $Y_{\mu}^{N} = \sum_{i=1}^{N} (\psi_{\mu}^{i})^{4}$.
The identity (\ref{IPRGreen}) 
holds only in the absence of degenerate states. In 
the presence of degenerate states, the function $\mathcal{G}^{N}_i(\lambda)$
is given by eq. (\ref{invPappaux}) plus a correction term
that involves a sum over all the degenerate eigenvectors.


\section{The cavity method}  \label{sec:appA}

We show in this appendix how to derive the 
cavity equations for the normalized complex
function $\mathcal{P}_{N,z}(\bx)$ defined by eq. (\ref{measure}). We 
focus here on the ensemble of fully-connected L\'evy matrices 
in which $J_{ii}=0$ for $\forall  \, i$. The values of the nondiagonal elements
of $\bJ$ are i.i.d.r.v.~drawn from the L\'evy distribution, defined
in the subsection \ref{seclevy}.

\subsection{Cavity equations}

The marginal at site $k$ is defined as follows
\begin{equation}
\mathcal{P}_{N,z}(x_{k}) = \int \left[ \prod_{j \in \partial_{k}} d x_{j} \right] \mathcal{P}_{N,z}(\bx) \,\,,
\label{marg}
\end{equation}
in which $\partial_{k}$ denotes the set of indices in a
row $k$ for which $J_{ij} \neq 0$. Here $\partial_{k}$ 
is composed of a number of indices of $O(N)$.

Using eq. (\ref{marg}) as a starting point, one can
derive the following equations
\begin{align}
\mathcal{P}_{N,z}(x_{k}) & \sim \int \left[ \prod_{j \in \partial_{k}} d x_{j} \right] 
\mathcal{P}^{(k)}_{N,z}(\bx) \nonumber \\
&\times \exp{\left( - \frac{i}{2} z  x^{2}_{k}  + i x_{k} \sum_{j \in \partial_{k}}^{N} J_{k j } x_{j} \right)} \,\,,
\label{probeqA} \\
\mathcal{P}^{(l)}_{N,z}(x_{k}) 
& \sim \int \left[ \prod_{j \in \partial_{k} \setminus l} d x_{j} \right] 
\mathcal{P}^{(k,l)}_{N,z}(\bx) \nonumber \\
&\times \exp{\left( - \frac{i}{2} z  x^{2}_{k}  + i x_{k} \sum_{j \in \partial_{k} 
\setminus l}^{N} J_{k j } x_{j} \right)} \,\,,
\label{probeq}
\end{align}
where $\partial_{k} \setminus l$ denotes the set $\partial_{k}$
without site $l$, and the function $\mathcal{P}^{(i_{1},\dots,i_{M})}_{N,z}(\bx)$ is defined on the
cavity graph $\mathcal{G}^{(i_{1},\dots,i_{M})}$. The cavity graph $\mathcal{G}^{(i_{1},\dots,i_{M})}$ is the
subgraph of the original graph $\mathcal{G}$, in which the nodes 
$i_{1},\dots,i_{M}$ and all their links with the other nodes have been removed.

In order to close the system of eqs. (\ref{probeqA}) 
and (\ref{probeq}), we make two assumptions which have
been used in the context of the cavity method for
disordered systems \cite{par1987}. First, we assume that the functions
$\mathcal{P}^{(i)}_{N,z}(\bx)$ and $\mathcal{P}^{(i,k)}_{N,z}(\bx)$
factorize over the sites according to $\mathcal{P}^{(i)}_{N,z}(\bx) = \prod_{j \in \partial_{i} }^{N} 
\mathcal{P}^{(i)}_{N,z}(x_j)$ and $\mathcal{P}^{(i,k)}_{N,z}(\bx) = \prod_{j \in \partial_{i} \setminus k  }^{N} 
\mathcal{P}^{(i,k)}_{N,z}(x_j)$, respectively.
Second, we assume that, in the limit $N \rightarrow \infty$, the 
marginals on the cavity graphs fulfill $\mathcal{P}^{(i)}_{N,z}(x_j) = 
\mathcal{P}^{(i,k)}_{N,z}(x_j)\,\, \forall \, j$.

When inserted in eqs. (\ref{probeqA}) and (\ref{probeq}), the above 
assumptions give rise to
\begin{align}
\mathcal{P}_{N,z}(x_{k}) & \sim  \exp{\left(- \frac{i}{2} z  x^{2}_{k} \right)} \nonumber \\
 &\times \prod_{j \in \partial_{k}}  \int d x_{j}  
 \mathcal{P}^{(k)}_{N,z}(x_j)  \exp{\left(  i x_{k} J_{k j } x_{j} \right)} 
\label{realP} \\
\mathcal{P}^{(l)}_{N,z}(x_{k}) & \sim 
\exp{\left(- \frac{i}{2} z  x^{2}_{k} \right)} \nonumber  \\
&\times  \prod_{j \in \partial_{k} \setminus l}  \int d x_{j}  
 \mathcal{P}^{(k)}_{N,z}(x_j)  \exp{\left(  i x_{k} J_{k j } x_{j} \right)} \,\,.
\label{cavP}
\end{align}
The form of eqs. (\ref{realP}) and (\ref{cavP}) suggest that they can be
solved in a self-consistent way through a Gaussian
assumption for the functions 
$\mathcal{P}^{(k)}_{N,z}(x_{i})$ and $\mathcal{P}_{N,z}(x_{i})$. The variances
of the local marginals $\mathcal{P}_{N,z}(x_{i})$ are the diagonal
elements of the Green function, as one can note
from eq. (\ref{Green1}). Thus we make the following
Gaussian {\it ansatz} \cite{RCKT08} for the
cavity functions $\mathcal{P}^{(l)}_{N,z}(x_{k})$ 
\begin{equation}
\mathcal{P}^{(l)}_{N,z}(x_{k}) = \sqrt{\frac{i}{2 \pi G^{N,(l)}_{k k}(z)}}
\exp{\left( -\frac{i x^{2}_{k}}{2 G^{N,(l)}_{k k}(z)} \right)} \,\,,
\label{Gausansatz}
\end{equation}
where $G^{N,(l)}_{k k}(z)$ are the diagonal elements of the Green function in which
row $l$ and column $l$ have been removed.
The substitution of the {\it ansatz} (\ref{Gausansatz}) in 
eqs. (\ref{realP}) and (\ref{cavP}) leads to the following 
self-consistent system of equations 
\begin{align}
G^{N,(k)}_{ii}(z) &= \frac{1}{z - g_{N,k}^{(i)}(z)   }\,, \label{caveqs} \\
G^{N}_{ii}(z) &= \frac{1}{z - h_{N}^{(i)}(z) } \,,
\label{realeqs}
\end{align}
for $i=1,\dots,N$ and for all $k \in \partial_{i}$. The 
functions $g_{N,k}^{(i)}(z)$ and $h_{N}^{(i)}(z)$ are defined as   
\begin{align}
g_{N,k}^{(i)}(z) &= \sum_{j \in \partial_{i} \setminus k} J^{2}_{i j} G^{N,(i)}_{j j}(z) \label{defg} \\
h_{N}^{(i)}(z) &= \sum_{j \in \partial_{i}} J^{2}_{i j} G^{N,(i)}_{j j}(z) \,\,. \label{defh}
\end{align}
The fixed-point solution of eqs. (\ref{caveqs}) and (\ref{realeqs})
allows one to determine the diagonal elements of the Green
function for a single instance of $\bJ$, which give access
to the DOS and the IPR. 


\subsection{The ensemble average}

In this subsection we explain how one can perform
the ensemble average and derive a self-consistent
equation for $W_{\lambda,\epsilon}(\omega)$ 
by employing a method introduced in \cite{NMB10, Jan10} for a fully-connected
L\'evy spin-glass.
The method consists in the introduction of
a small cutoff $\gamma$ that makes a distinction between small
and large matrix elements. The global contribution of the small
matrix elements is taken into account by means of
the law of large numbers.

We define the sets of indices that distinguish between
weak and strong matrix elements in a certain row $i$ 
according to
\begin{align}
\zeta_{i}(\gamma) &= \{  j  \in \mathbb{N} \cap  [1,N]|(J_{i j} < \gamma) \wedge (j \neq i)  \}\,\,, \nonumber \\
\overline{\zeta}_{i}(\gamma) &= \{ j  \in \mathbb{N} \cap  [1,N]|   (J_{i j} > \gamma) \wedge (j \neq i)  \} \,\,. \nonumber
\end{align}
These definitions allow us to rewrite
$g_{N,k}^{(i)}(z)$ and $h_{N}^{(i)}(z)$ as follows
\begin{align}
h_{N}^{(i)}(z) &= \sum_{j \in \overline{\zeta}_{i}(\gamma) } J^{2}_{i j} G^{N,(i)}_{j j}(z) 
+ \sum_{j \in \zeta_{i}(\gamma)} J^{2}_{i j}  G^{N,(i)}_{j j}(z)  \nonumber \\
k \in \overline{\zeta}_{i}(\gamma): \nonumber \\
g_{N,k}^{(i)}(z) &=  \sum_{j \in \overline{\zeta}_{i}(\gamma) \setminus k} J^{2}_{i j} G^{N,(i)}_{j j}(z) 
+ \sum_{j \in \zeta_{i}(\gamma)} J^{2}_{i j} G^{N,(i)}_{j j}(z)  \nonumber \\
k \in \zeta_{i}(\gamma):\nonumber \\
g_{N,k}^{(i)}(z) &= \sum_{j \in \overline{\zeta}_{i}(\gamma) } J^{2}_{i j} G^{N,(i)}_{j j}(z) 
+ \sum_{j \in \zeta_{i}(\gamma)\setminus k} J^{2}_{i j} G^{N,(i)}_{j j}(z)  \nonumber
 \,.
\end{align}
In the limit $N \rightarrow \infty$, we can
remove the $k$ dependence from the sum over
the weak matrix elements because it contains an infinite
number of terms. Defining the joint distribution $\Omega^{(j)}_{z}(\omega)$ 
of the real and imaginary parts of $G^{N,(j)}_{i i}(z)$ for a
fixed $j$
\begin{equation}
\Omega^{(j)}_{z}(\omega) = \lim_{N \rightarrow \infty} \frac{1}{N} \sum_{i=1}^{N}      
\delta\left[ \omega - G^{N,(j)}_{i i}(z) \right] \,\,,
\label{distrcav}
\end{equation}
one can apply the law of large numbers to the 
contribution coming from the weak matrix elements, leading to
\begin{equation}
\lim_{N \rightarrow \infty} \sum_{j \in \zeta_{i}(\gamma)} J^{2}_{i j} G^{N,(i)}_{j j}(z)  
= \sigma_{\gamma}^{2} \int d \omega \Omega^{(i)}_{z}(\omega) \omega \,\,,
\end{equation}
where $\sigma_{\gamma}^{2}$ is the variance of the weak matrix 
elements \cite{NMB10}
\begin{equation}
\sigma_{\gamma}^{2} = N \int_{-\gamma}^{\gamma} dJ P_{\alpha}(J) J^2 = 
\frac{2 \gamma^{2-\alpha} C_{\alpha}}{2-\alpha} \,\,,
\end{equation}
with the distribution $P_{\alpha}(J)$ defined by eqs.~(\ref{defStable}) and 
(\ref{defChar}). 

A self-consistent equation for $\Omega^{(j)}_{z}(\omega)$ is derived
by substituting eq. (\ref{caveqs}) in (\ref{distrcav}). One has
to distinguish between two cases. If $j \in \zeta_{i}(\gamma)$,
the function $g_{N,j}^{(i)}(z)$ is equal to $h_{N}^{(i)}(z)$ for
any index $i$, and as a consequence one obtains $\Omega^{(j)}_{z}(\omega) = 
W_{\lambda,\epsilon}(\omega)$. If $j \in \overline{\zeta}_{i}(\gamma)$, one
can follow the discussion of \cite{NMB10} in order
to show that $\Omega^{(j)}_{z}(\omega) = W_{\lambda,\epsilon}(\omega)$. The
distribution $W_{\lambda,\epsilon}(\omega)$ fulfills the self-consistent
equation
\begin{align}
&W_{\lambda,\epsilon, \gamma}(\omega) = \sum_{k=0}^{\infty} \frac{e^{-c_{\gamma}} \, c_{\gamma}^{k}}{k!}
\int \left[ \prod_{l=1}^{k} d \omega_{l} W_{\lambda,\epsilon, \gamma}(\omega_{l}) \right] \nonumber \\
&\times \int \left[ \prod_{l=1}^{k} d J_{l} \, p_{J, \gamma}(J_{l}) \right] \nonumber \\
&\times \delta \left(\omega - \frac{1}{z - \sigma_{\gamma}^{2} 
\langle \omega \rangle -  \sum_{l=1}^{k} H(\omega_{l},J_{l})} \right) \,\,,
\label{eqWLevyap}
\end{align}
in which $\langle f(\omega) \rangle = \int d \omega W_{\lambda,\epsilon, \gamma}(\omega) f(\omega)$
and
\begin{align}
H(\omega,J) &= J^{2} \omega \,\,,   \\
c_{\gamma} &= \frac{2 C_{\alpha}}{\alpha \gamma^{\alpha}} \,\,, \\
p_{J,\gamma}(J) &= \left\{ \begin{array}{ccc}\frac{\alpha \gamma^{\alpha}}
{2 |J|^{\alpha+1}}&  &|J|>\gamma \\ 0& &|J|<\gamma  \end{array}
\right. \,\,.
\end{align}
The quantity $C_{\alpha}$ is defined by eq. (\ref{Cdef}). 


\acknowledgments

We would like to thank Andrea Pagnani for a helpful discussion. 
IN thanks Isaac P\'erez Castillo and Tim Rogers for many interesting discussions.
FLM thanks Yan Fyodorov for a useful correspondence.

\bibliography{bibliography}

\end{document}